\documentclass[aps,prl,twocolumn, superscriptaddress, showkeys,longbibliography]{revtex4-2}  

\usepackage{graphicx}% Include figure files
\usepackage{dcolumn}% Align table columns on decimal point
\usepackage{bm}% bold math
\usepackage{xcolor}
\usepackage{ulem} 
\usepackage{amsmath}
\allowdisplaybreaks
\usepackage[bookmarks,
                bookmarksopen = true,
                bookmarksnumbered = true,
                linktocpage,
                colorlinks = true,
                linkcolor = blue,
                urlcolor  = blue,
                citecolor = blue,
                anchorcolor = green,
                hyperindex = true,
                hyperfigures]
                {hyperref}
\usepackage{multirow}
\usepackage{array}
\usepackage{epstopdf}
\usepackage{upgreek}

\newcommand{\sqrtsnn}{\ensuremath{\sqrt{s_{\mathrm {NN}}}}}

\newcommand{\IAA}{\ensuremath{I_\mathrm{AA}}}

\begin{document}

\title{Emergence of thermal recoil jets in high-energy heavy-ion collisions}

\author{Peng Jing}
\affiliation{Institute of Frontier and Interdisciplinary Science, Shandong University, Qingdao, Shandong 266237, China}

\author{Yichao Dang}
\affiliation{Institute of Frontier and Interdisciplinary Science, Shandong University, Qingdao, Shandong 266237, China}

\author{Lejing Zhang}
\affiliation{Institute of Frontier and Interdisciplinary Science, Shandong University, Qingdao, Shandong 266237, China}

\author{Yang He}
\affiliation{Department of Modern Physics, University of Science and Technology of China, Hefei, Anhui 230026, China}

\author{Shanshan Cao}
\email{shanshan.cao@seu.edu.cn}
\affiliation{School of Physics, Southeast University, Nanjing, Jiangsu 211189, China}
\affiliation{Institute of Frontier and Interdisciplinary Science, Shandong University, Qingdao, Shandong 266237, China}

\author{Li Yi}
\email{li.yi@sdu.edu.cn}
\affiliation{Institute of Frontier and Interdisciplinary Science, Shandong University, Qingdao, Shandong 266237, China}

\author{Xin-Nian Wang}
\email{xnwang@ccnu.edu.cn}
\affiliation{Institute of Particle Physics and Key Laboratory of Quark and Lepton Physics (MOE), Central China Normal University, Wuhan, Hubei 430079, China}

\date{\today}

\begin{abstract}

In the established paradigm of jet quenching in relativistic heavy-ion collisions, jets from initial hard parton scatterings are suppressed due to their interaction with the quark-gluon plasma (QGP), serving as crucial tomographic probes of QGP properties. Within the linear Boltzmann transport model, we find that the QGP is also capable of absorbing and reprocessing energy deposited by the hard jets into emergent jet-like objects, providing an alternative production mechanism of thermal recoil jets. These emergent thermal recoil jets exhibit distinct transverse momentum ($p_\mathrm{T}$) and jet-cone size ($R$) dependencies different from the hard jets, and interpret the puzzling observation of the enhanced yields of hadron triggered jets at large azimuthal angle relative to the away side and solely at small $p_\mathrm{T}$ and large $R$. These thermal recoil jets are predicted to have unique substructures, such as a jet shape that increases with radius and a thermal-like distribution of their constituents, which await verification in future experimental analyses.

\end{abstract}

\maketitle

{\it \color{blue} Introduction -- } Creating and studying quark-gluon plasma (QGP), a deconfined state of nuclear matter consisting of quarks and gluons, in relativistic heavy-ion collisions is a central focus of high-energy nuclear physics~\cite{Gyulassy:2004zy,Jacobs:2004qv, Busza:2018rrf,Elfner:2022iae,Harris:2024aov}. Experimental data on jet quenching at the Relativistic Heavy Ion Collider (RHIC) and the Large Hadron Collider (LHC), a phenomenon caused by the energy loss of high energy partons as they traverse the QGP, have provided important information about the interaction between jets and the hot QGP medium~\cite{Wang:1992qdg,Qin:2015srf,Majumder:2010qh,Cao:2020wlm,Wang:2025lct,Mehtar-Tani:2025rty,JET:2013cls,JETSCAPE:2021ehl}. Jet quenching can manifest as a reduction of jet yields triggered by high-transverse-momentum ($p_\mathrm{T}$) particles~\cite{ALICE:2015mdb,ATLAS:2018dgb,CMS:2012ytf,CMS:2017eqd,STAR:2017hhs,STAR:2023ksv,STAR:2023pal} where the high-$p_\mathrm{T}$ particles, in particular $\gamma$ or $Z$ bosons, are used as reference of the initial hard scattering scale. Using a statistical approach to correct for uncorrelated background, the semi-inclusive coincidence measurements of hadron and photon triggered jets ($h$-jets and $\gamma$-jets) have been able to access a broader range of jet kinematics, especially at low jet $p_\mathrm{T}$ and with large jet-cone size ($R$)~\cite{ALICE:2015mdb,ALICE:2015mdb,ALICE:2023qve,ALICE:2023jye,STAR:2017hhs,STAR:2023ksv,STAR:2023pal,STAR:2025yhg}.
Examining the nuclear modification of these low $p_\mathrm{T}$ and large $R$ jets recoiling from high-$p_\mathrm{T}$ triggers may uncover the dynamics of jet production and evolution, as well as the medium response that is sensitive to the microscopic structures of the QGP. 

In the study of hadron-jet correlations, the ALICE Collaboration reported the first observation of an enhancement of recoil jet yield and acoplanarity in central Pb+Pb collisions at $\sqrt{s_\mathrm{NN}} = 5.02$~TeV~\cite{ALICE:2023qve,ALICE:2023jye}. A $4.7\sigma$ signal of enhanced jet yields up to a factor of 6 at large azimuthal deviations from $\Delta\phi = \pi$ relative to the trigger hadron is observed at $10 < p_\mathrm{T,jet} < 20$~GeV for jets with cone-size $R = 0.4$, $0.5$, but not for $R=0.2$. Similar $R$-dependence of the enhancement of photon and pion triggered recoil jets has also been observed by the STAR Collaboration in central Au+Au collisions at $\sqrt{s_\mathrm{NN}} = 200$~GeV~\cite{STAR:2025yhg}. 
These observations contradict the expected picture of jet deflection via Moli\`ere scattering with quasi-particles inside the QGP, whose enhanced jet yields at large angles should not obviously depend on the jet-cone size. Current theoretical calculations on $h$-jets and $\gamma$-jets face critical challenges. So far, no model in the literature is capable of quantitatively describing the $p_\mathrm{T,jet}$, $R$, and $\Delta \phi$ dependencies of the recoil jet yield simultaneously. These include both perturbative calculations at next-to-leading-log~\cite{Chen:2016vem} and several sophisticated Monte-Carlo event generators like JETSCAPE~\cite{JETSCAPE:2021ehl}, JEWEL~\cite{Zapp:2008gi,Zapp:2013vla}, and the Hybrid model~\cite{Casalderrey-Solana:2014bpa}, which have successfully described other aspects of jet quenching in heavy-ion collisions. 

Within our model, we will show that the large enhancement of recoil jets at $\Delta \phi$ away from $\pi$ relative to the trigger is not driven by medium modification of ``hard jets" that originate from initial parton scatterings. Instead, 
jet-induced medium response can give rise to jet-like objects from the QGP medium, resulting in the observed enhancement of low-$p_\mathrm{T}$ jet yields. Jet-induced medium response~\cite{Cao:2022odi,Yang:2025lii}, caused by the energy deposited into the medium by hard jets, is an essential part of jet-medium interaction that affects nearly all observables of full jets, from the enhancement of the nuclear modification factor to the elliptic flow coefficient of jets~\cite{He:2018xjv,He:2022evt}, the transverse energy distribution (jet shape) at large radius relative to the jet axis~\cite{Tachibana:2017syd,Casalderrey-Solana:2016jvj,KunnawalkamElayavalli:2017hxo,Luo:2018pto,Ke:2020clc}, the longitudinal fractional momentum ($z$) distribution (fragmentation function) at small $z$~\cite{KunnawalkamElayavalli:2017hxo,Chen:2017zte,Chen:2020tbl,Ke:2020clc}, jet mass~\cite{KunnawalkamElayavalli:2017hxo,Park:2018acg,Casalderrey-Solana:2019ubu,Luo:2019lgx}, and energy-energy correlators at large relative angles~\cite{Yang:2023dwc,Xing:2024yrb}. Unique features of medium response, including energy depletion in the direction opposite to jet propagation~\cite{Chen:2017zte,Yang:2021qtl,Yang:2022nei,Yang:2025dqu} and enhanced strangeness and baryon production inside medium-modified jets~\cite{Chen:2021rrp,Luo:2021voy,Sirimanna:2022zje,Luo:2024xog}, have been proposed and are currently under active experimental search~\cite{ATLAS:2024prm,CMS:2025dua,Dale-Gau:2025nyw,Cantway:2025yao}.  
We consider in this Letter the possibility of creating jets from the medium response in the QGP background. We will demonstrate that thermal recoil particles in jet-medium scatterings can indeed produce jet-like objects, which contribute to the observed enhancement of the yield and acoplanarity of low-$p_{\rm T}$ and large cone-size $h$-jets within our model. These jet-like objects have features distinct from hard jets produced by initial parton scatterings in observables such as the jet shape and fragmentation function. We refer to these jet-like objects as ``thermal recoil jets".

{\it \color{blue} Medium modification of $h$-jets -- }
We use the linear Boltzmann transport (LBT) model~\cite{Wang:2013cia,He:2015pra,Cao:2016gvr,Xing:2019xae,Luo:2023nsi,Dang:2026ezw} based on the Boltzmann equation,
\begin{equation} 
	p_a \cdot \partial f_a = E_a\left[C^{\rm el}(f_a) + C^{\rm inel}(f_a\right)],
\end{equation}
to simulate the evolution of the phase space distribution $f_a(t,\vec{x}_a,\vec{p}_a)$ of a parton with four-momentum $p_a=(E_a,\vec{p}_a)$, which can be a jet parton, thermal recoil parton or radiated gluon from the jet-medium interaction. The collision kernel for elastic scatterings $C^{\rm el}$ is evaluated using the matrix elements of $ab\to cd$ processes with summation over all possible $b$, $c$, and $d$. The collision kernel for inelastic scatterings, $C^{\rm inel}$, is calculated from the medium-induced gluon spectrum within the higher-twist approach to parton energy loss~\cite{Wang:2001ifa,Zhang:2003wk,Majumder:2009ge}. Pythia~8~\cite{Sjostrand:2014zea,Sjostrand:2006za} is used to generate hard jet partons from the initial parton scatterings in nucleon-nucleon collisions. These partons undergo splittings (vacuum showering) in Pythia until the virtuality scale of each daughter parton reaches the thermal scale of the QGP (set as $Q_\mathrm{M}=2$~GeV). These partons stream freely from the hard collision vertices, which are sampled using the Monte-Carlo model~\cite{Miller:2007ri} for nucleus-nucleus (A$+$A) collisions, for the duration of their formation time ($\tau_\mathrm{form}$---sum of their previous splitting times)~\cite{Zhang:2022ctd} before they propagate and evolve in the QGP medium according to the Boltzmann equation.

Evolution of the QGP medium is described by the (3+1)-dimensional viscous hydrodynamic model CLVisc~\cite{Pang:2012he,Pang:2018zzo,Wu:2018cpc,Wu:2021fjf} that is constrained by the experimental data on low-$p_\mathrm{T}$ hadron spectra in heavy-ion collisions. Each parton from the initial hard jets starts interacting with the medium at $t_\mathrm{init}=\max(\tau_0,\tau_\mathrm{form})$, with $\tau_0=0.6$~fm the starting time of the hydrodynamic evolution. Collision kernels in the Boltzmann equation are evaluated at the local temperature and flow velocity of the medium provided by hydrodynamic simulations with the initial energy density distribution from the Trento model~\cite{Moreland:2014oya}.  The LBT model keeps track of all partons involved in the parton-medium interaction: jet partons, thermal recoil partons, radiated gluons as well as ``negative partons'' from the back-reaction which have to be subtracted from the final parton phase space distributions.  Recoil partons and radiated gluons are allowed to further scatter with the QGP in the same way as for jet partons. Recoil and negative partons constitute medium response in our study. The virtuality scales of the final jet partons from Pythia are kept as $Q_\mathrm{M}$ in LBT, while the scales of recoil and negative partons are set as the lower limit in Pythia, or hadronization scale $Q_\mathrm{h}=0.5$~GeV.

\begin{figure}[tbp!]
    \centering
    \includegraphics[width=0.85\linewidth]{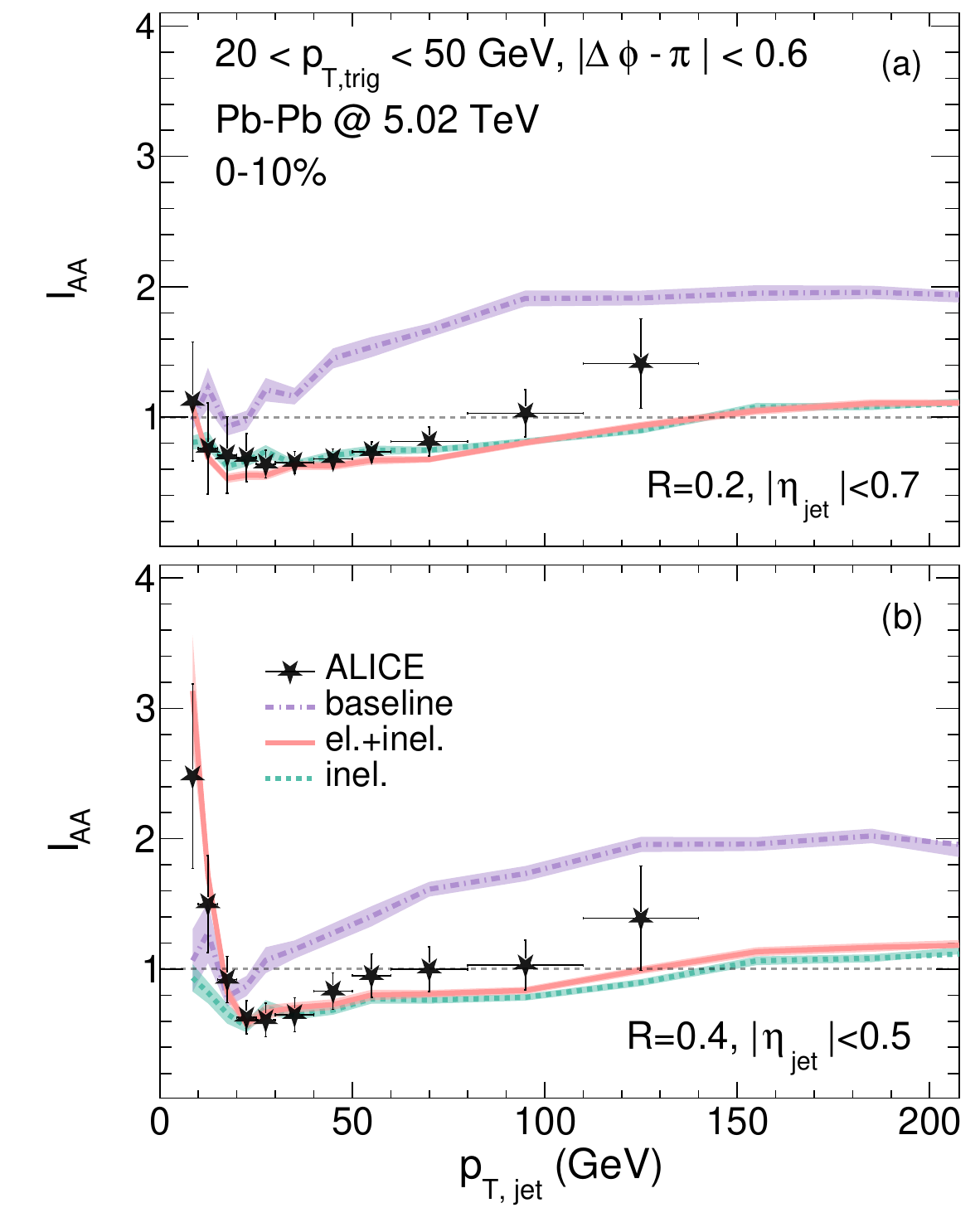}
    \caption{(Color online) The $\IAA$ of hadron triggered charged jets as a function of the jet $p_\mathrm{T}$ in central (0-10\%) Pb+Pb collisions at $\sqrtsnn = 5.02$~TeV from LBT simulations for unquenched jets (baseline, dot-dashed), quenched jets with only inelastic scatterings (dashed) and with both elastic and inelastic scatterings (solid), compared to the ALICE data~\cite{ALICE:2023qve}, (a) for jets with radius $R=0.2$ and (b) for $R=0.4$.}
    \label{fig:1}
\end{figure}

When a parton exits the QGP boundary, defined by the hadronization temperature of the QGP ($T_\mathrm{h}=165$~MeV), it is fed back to Pythia for its subsequent vacuum showering down to its hadronization scale $Q_\mathrm{h}$, after which it is converted to hadrons via string fragmentation in Pythia. Regular and negative partons are converted to hadrons separately, and the contribution from the latter is subtracted from that of the former for all jet observables. The final hadrons are clustered into jets using a modified FastJet package~\cite{Cacciari:2011ma} where the momenta of negative hadrons are subtracted from those of regular ones~\cite{He:2018xjv}.

In this work, we use a recently updated version of the LBT model~\cite{Dang:2026ezw,githubLBT} that tracks the color flows of partons as they evolve through the QGP. This allows one to execute jet-medium interactions at $Q_\mathrm{M}>Q_\mathrm{h}$ and then pass partons back to Pythia for further vacuum showers and string hadronization. Non-perturbative effects on jet evolution are modeled by enhancing the perturbative-based jet transport coefficient at low momentum with a $K$ factor, which is constrained by heavy quark data at low $p_\mathrm{T}$~\cite{Cao:2016gvr}. This model simultaneously describes the nuclear modification factors of jets and hadrons with different flavors. The same model setting is used for all calculations in the present work.

\begin{figure}[tbp!]
    \centering
    \includegraphics[width=0.95\linewidth]{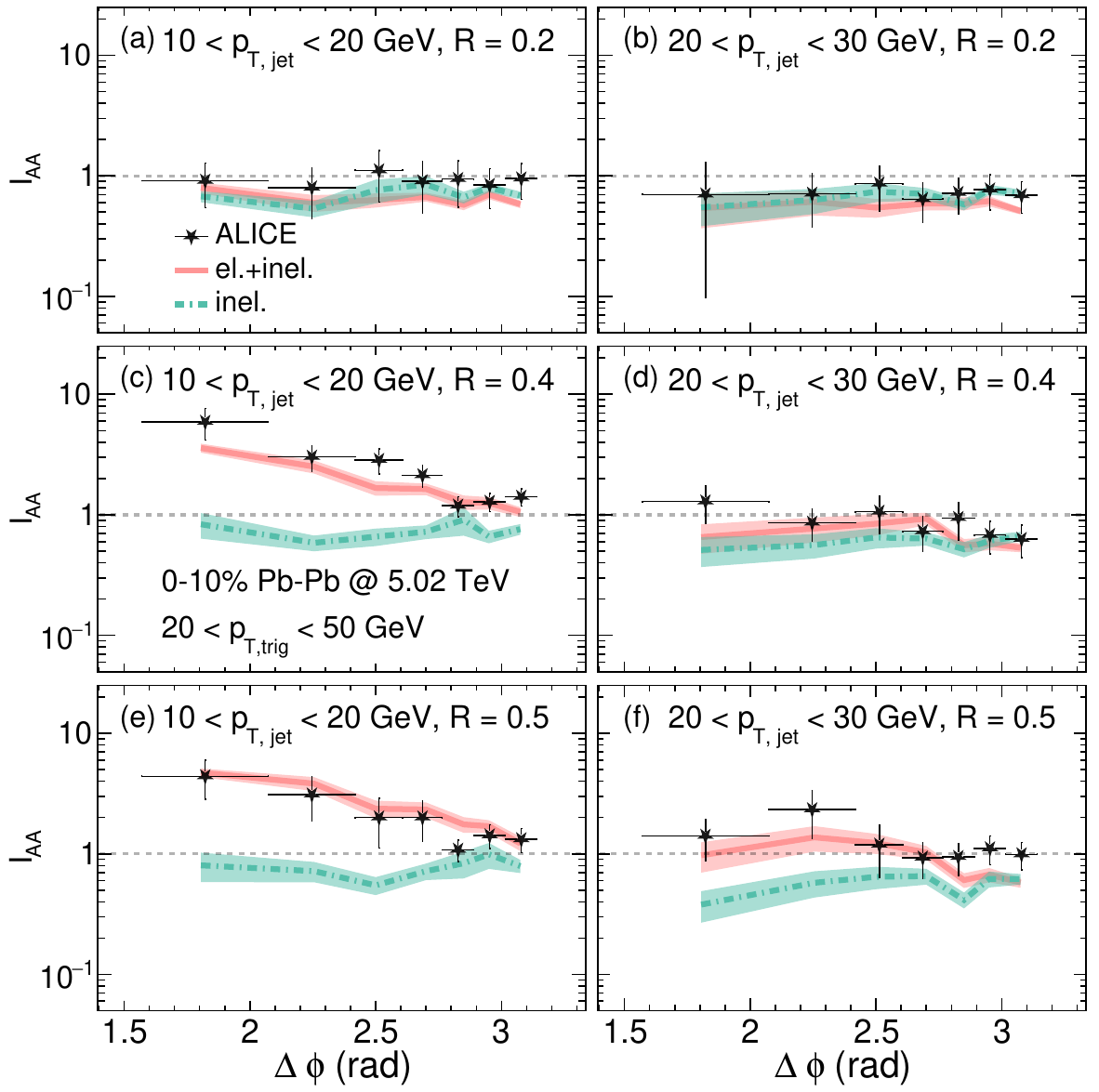}
    \caption{(Color online) The $\IAA$ of $h$-jets as a function of the azimuthal angle between jets and their trigger hadrons in central (0-10\%) Pb+Pb collisions at $\sqrtsnn = 5.02$~TeV, compared between the LBT calculations with and without including elastic scatterings, and the ALICE data~\cite{ALICE:2023qve}. Different panels are for different $p_\mathrm{T,jet}$ and $R$ of the recoil jets.}
    \label{fig:2}
\end{figure}

The medium modification of $h$-jets is quantified by the $\IAA$ factor defined as the ratio of $h$-jet spectrum in A$+$A collisions to that in $p+p$. Plotted in Fig.~\ref{fig:1} are our results on the $\IAA$ of hadron triggered charged jets as functions of jet $p_\mathrm{T,jet}$ in 0-10\% Pb+Pb collisions at $\sqrtsnn = 5.02~\mathrm{TeV}$ with two different jet-cone sizes $R=0.2$ [panel (a)] and $R=0.4$ [panel (b)]. Here, trigger hadrons are selected with $20 < p_\mathrm{T,trig} < 50$~GeV and recoil jets are restricted to the azimuthal region of $|\Delta \phi - \pi| < 0.6$ relative to the trigger according to the selection in ALICE measurements~\cite{ALICE:2023qve}. To help understand the behavior of $\IAA$, we also show the baseline case (dot-dashed lines) where the trigger hadron comes from a quenched jet while the associated jet does not interact with the medium and therefore is not modified. Due to energy loss, the initial energy of the jet that produces the trigger hadron for a given value $p_\mathrm{T,trig}$ in A$+$A collisions is larger than that in $p+p$ collisions, causing $\IAA>1$ in the baseline case. This is known as the trigger bias that leads to the observed $\IAA>1$ in both the baseline and the full calculation (solid lines) at high $p_\mathrm{T,jet}$ even after taking into account the medium modification of the recoil jets~\cite{He:2024rcv}. Such trigger bias does not exist for $\gamma$-jets. An enhancement of $\IAA$ above 1 is also observed in panel (b) at low $p_\mathrm{T,jet}$ for jets with $R=0.4$. This enhancement could result from medium response, which indeed disappears in our model calculation (dotted line) when elastic process is switched off. Elastic scattering is a primary mechanism for generating recoil particles in LBT though inelastic processes also produce thermal recoil partons in addition to radiated gluons. Since elastic scattering also contributes to jet energy loss and momentum broadening, a firm conclusion about the original of the $I_{\rm AA}$ enhancement of large-$R$ $h$-jets at low $p_\mathrm{T,jet}$ requires isolating the contribution from medium response at the partonic level as we show in the next section.

To examine the $h$-jet modification in the low $p_{\rm T,jet}$ region in detail, we show in Fig.~\ref{fig:2} the azimuthal angle dependence of the $h$-jet $\IAA$ in two ranges of $p_{\rm T,jet}$ with three different jet cone sizes associated with a trigger hadron with $20 < p_\mathrm{T,trig} < 50$~GeV. In panels (c) and (e), one observes a large enhancement of $\IAA$ at large $\Delta\phi-\pi$ for recoil jets with $10<p_\mathrm{T,jet}<20$~GeV and $R=0.4$ or $0.5$. This enhancement vanishes for $h$-jets with small cone-size $R=0.2$ [panels (a) and (b)], and is significantly reduced for large $p_\mathrm{T,jet}$ [panels (d) and (f)]. The large enhancement disappears when elastic scattering is disabled in LBT (dot-dashed), indicating that medium response could be responsible for the enhanced low-$p_{\rm T,jet}$ and large-$R$ $h$-jets in LBT. With both elastic and inelastic scatterings activated (solid), the LBT model provides a simultaneous description of the $p_\mathrm{T,jet}$ and $\Delta \phi$ dependencies of the $h$-jet $\IAA$ in different kinematic regions and with different jet cone-sizes. The qualitative conclusion that the $\IAA$ enhancement observed for low-$p_{\rm T,jet}$ and large-$R$ jets likely results from medium response is also supported by the Hybrid and JEWEL model calculations as shown in Ref.~\cite{ALICE:2023qve}. The $K$ factor and the hadronization scheme implemented in the improved LBT model are fixed by comparisons to the nuclear modifications factors of inclusive hadrons and jets~\cite{Dang:2026ezw}. No additional tuning is required for the $\IAA$ calculation in this study. Without these, the $\IAA$ would be reduced to around or below 2 at large $\pi-\Delta\phi$ in Fig.~\ref{fig:2} (c) and (e).

\begin{figure}
    \centering
    \includegraphics[width=0.85\linewidth]{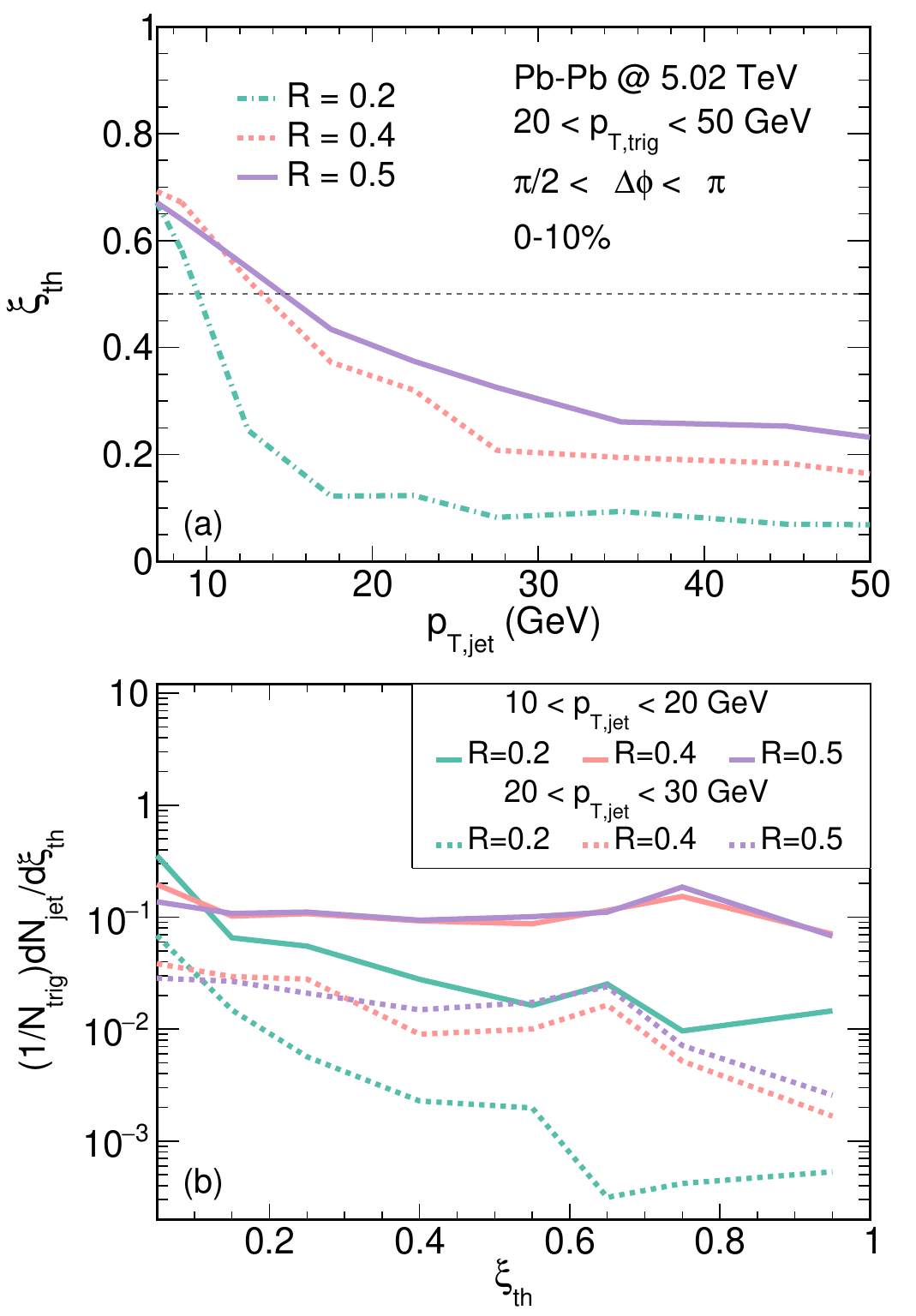}
    \caption{(Color online) The fractional $p_\mathrm{T}$ of recoil partonic jets contributed by the medium response ($\xi_\mathrm{th}$) in central (0-10\%) Pb+Pb collisions at $\sqrtsnn=5.02$~TeV. Panel (a) compares the average values of $\xi_\mathrm{th}$ between different jet radii, as functions of $p_\mathrm{T,jet}$; Panel (b) compares the $\xi_\mathrm{th}$ distributions between different $p_\mathrm{T,jet}$ and jet radii.}
    \label{fig:3}
\end{figure}

{\it \color{blue} Emergence of thermal recoil jets -- } Since partons from different sources are combined in string hadronization, we investigate jet components at the partonic level in order to theoretically identify the mechanism underlying the $\IAA$ enhancement at large $\pi-\Delta\phi$. For parton-triggered partonic jets, we define the fractional transverse momentum of a jet contributed by medium response as,
\begin{equation}
    \xi_{\rm th} = \frac{|\vec{p}_\mathrm{T}^{\;\mathrm{recoil}} - \vec{p}_\mathrm{T}^{\;\mathrm{negative}}|}{|\vec{p}_\mathrm{T}^{\;\mathrm{recoil}} - \vec{p}_\mathrm{T}^{\;\mathrm{negative}}| + p_\mathrm{T}^{\text{jet-shower}}},
\end{equation}
where the ``jet-shower" part contains contributions from the final states of the Pythia jet shower partons and their radiated gluons, and ``recoil" contains contributions from the offspring of the recoil partons. 

As shown in Fig.~\ref{fig:3}(a), for recoil jets within ${\pi}/{2}<\Delta\phi<\pi$ triggered by partons with $20 < p_\mathrm{T,trig} < 50$~GeV, the average momentum fraction $\xi_{\rm th}$ from medium response inside the recoil jets can exceed 0.5 when $p_\mathrm{T,jet}$ is low. At a given $p_\mathrm{T,jet}$, $\xi_{\rm th}$ is generally higher for jets with larger $R$. The $\xi_{\rm th}$ distributions of recoil jets within different kinematic ranges are presented in Fig.~3(b), where jets with small cone-size $R$ or large $p_\mathrm{T,jet}$ are distributed mostly at small $\xi_{\rm th}$. However, a significant fraction (over 49\%) of the low-$p_\mathrm{T,jet}$ and large-$R$ jets have $\xi_{\rm th}>0.5$, and therefore should no longer be simply viewed as medium-modified hard jets originating from initial hard partonic scatterings. Since medium response particles (recoil and negative particles in LBT) are excited from the thermal QGP background, we call these jets with large $\xi_{\rm th}$ ``thermal recoil jets". We have verified that the average value of $\xi_{\rm th}$ is not sensitive to $\Delta \phi$. Given that the recoil jet yield in $p+p$ collisions drops rapidly as $\pi-\Delta\phi$ becomes larger, a considerable number of thermal recoil jets with low $p_\mathrm{T,jet}$ and large $R$ in A$+$A collisions lead to the enhancement of $\IAA$ at large $\pi-\Delta\phi$. This also interprets why such enhancement disappears for jets with large $p_\mathrm{T,jet}$ or small $R$ as seen in Fig.~\ref{fig:2} within the LBT model. Note that this $\xi_{\rm th}$ value is a parton-level quantity computed within the LBT model and is not an experimental observable for jet classification.

\begin{figure}[tbp!]
    \centering
    \vspace{-5pt}
    \includegraphics[width=0.8\linewidth]{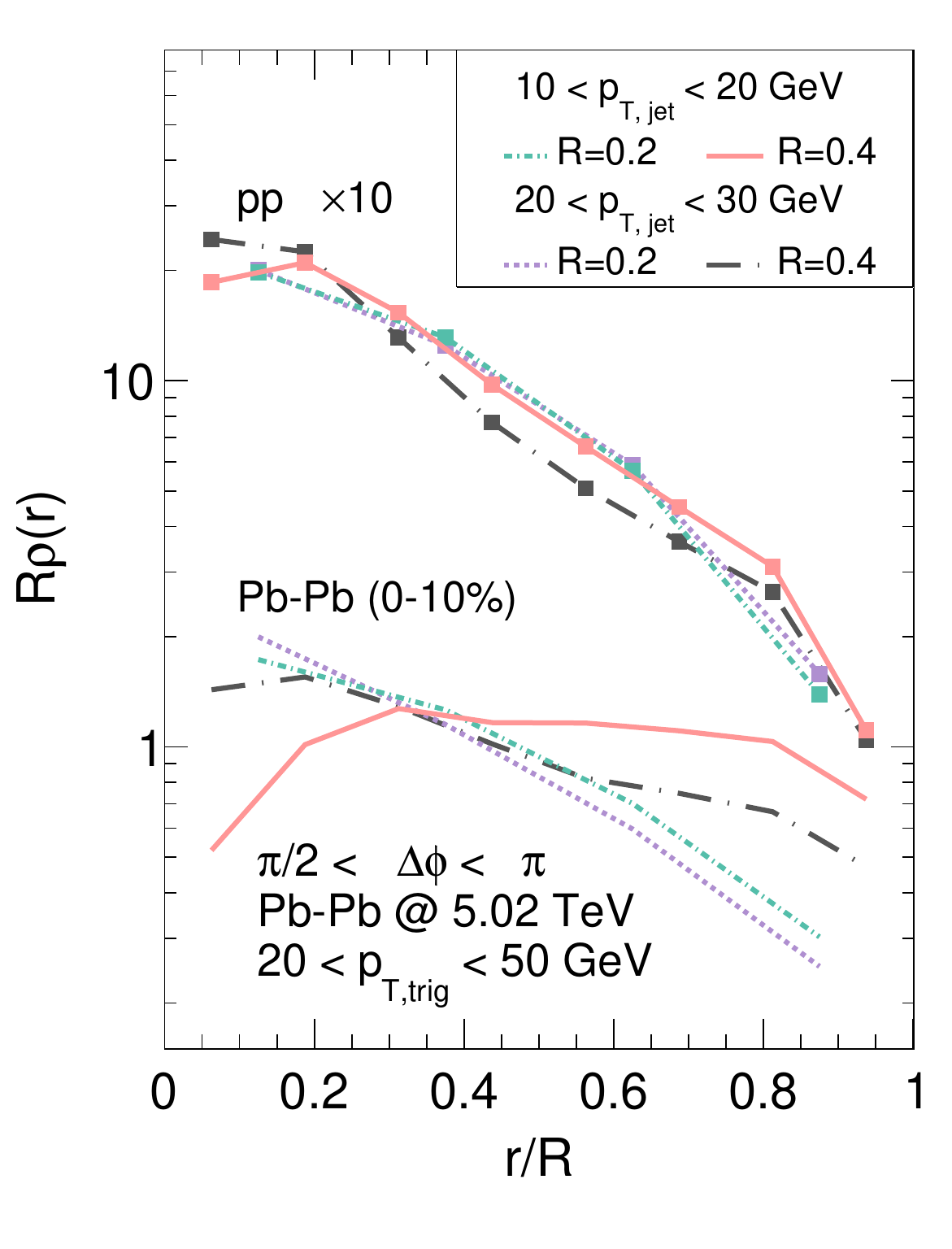}
    \vspace{-8pt}
    \caption{(Color online) The jet shape of hadron triggered charged jets $R\rho(r)$ as a function of the scaled radius $r/R$ in $p+p$ and central Pb+Pb collisions at $\sqrtsnn = 5.02~\mathrm{TeV}$ in different $p_\mathrm{T,jet}$ ranges and with different jet-cone size $R$. 
    }
    \label{fig:4}
\end{figure}

To experimentally verify the thermal nature of these thermal recoil jets, one may analyze the transverse energy distribution inside jets, or jet shape, defined as~\cite{CMS:2020geg}
\begin{equation}
\rho(r) = \frac{1}{\delta r} \frac{\sum_{\mathrm{jets}}\sum_{\mathrm{track} \in \left( r - \frac{\delta r}{2},\ r + \frac{\delta r}{2} \right)} p_\mathrm{T, track}}{\sum_{\mathrm{jets}}\sum_{\mathrm{track}} p_\mathrm{T, track}},
\end{equation}
with $r$ the radial distance to the jet axis, $\delta r$ the width of an annular ring, $p_\mathrm{T,track}$ the $p_\mathrm{T}$ carried by particles. In Fig.~\ref{fig:4}, we show jet shapes of hadron triggered charged jets in $p+p$ and central (0-10\%) Pb+Pb collisions at $\sqrtsnn=5.02$~TeV as a function of the scaled radius $r/R$. For each collision system, the four curves correspond to the upper four panels of Fig.~\ref{fig:2} for different $p_\mathrm{T,jet}$ and $R$. For jets with $R=0.2$ in Pb+Pb collisions, one observes a fast decrease of $\rho(r)$ as $r$ increases. This is in qualitative agreement with the jet shape observed in $p+p$ collisions, and is the typical shape of hard and collimated jets in which energetic particles carry most of the energy tightly around the core while soft particles carry small fractions of the energy at large $r$. In contrast, the jet shape $\rho$ increases with $r$ first and then becomes flat for jets with $R=0.4$ and $10<p_\mathrm{T,jet}<20$~GeV in Pb+Pb collisions. This is the kinematic range where thermal recoil jets have a large population and a significant enhancement of $\IAA$ appears at $\Delta \phi$ away from $\pi$ in Fig.~\ref{fig:2}. Such a particular jet shape results from the sparsely distributed soft particles from medium response in the momentum space, which also helps understand why a large $R$ is necessary to cluster sufficient energy into a cone to form a thermal recoil jet which usually carries low transverse momenta. This qualitatively different jet shape can be used to identify thermal recoil jets in heavy-ion experiments. 

\begin{figure}
    \centering
    \vspace{-5pt}
    \includegraphics[width=0.8\linewidth]{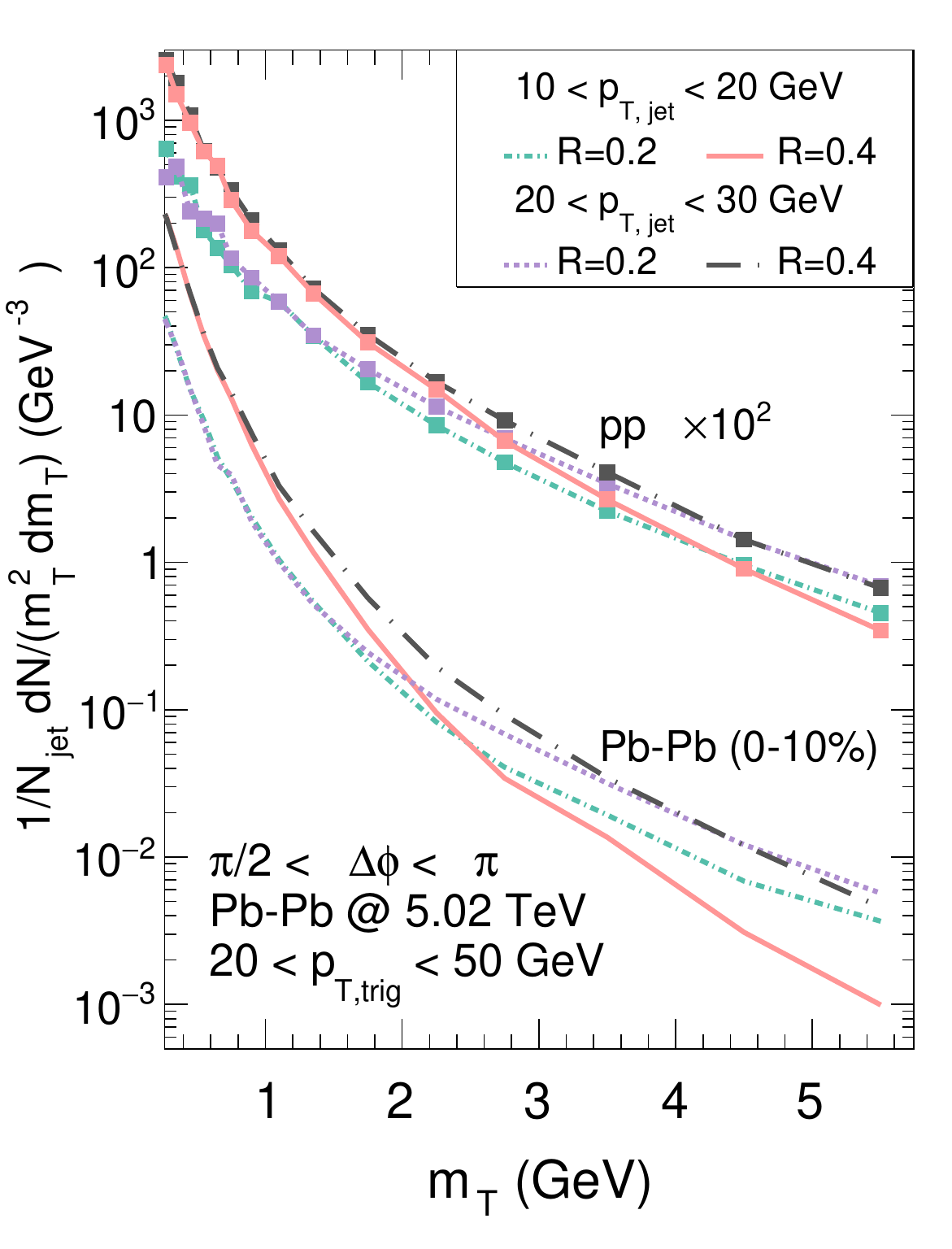}
    \vspace{-8pt}
    \caption{(Color online) The transverse mass distribution of pions inside the recoil charged jets in $p+p$ and central Pb+Pb collisions at $\sqrtsnn = 5.02~\mathrm{TeV}$ in different $p_\mathrm{T,jet}$ ranges and with different jet-cone size $R$.
    }
    \label{fig:5}
\end{figure}

Another signature of thermal recoil jets could be the thermal-like distribution of particles inside these jets. Shown in Fig.~\ref{fig:5} is the transverse mass distribution---$ 1/N_\mathrm{jet} [dN / (m_\mathrm{T}^2 dm_\mathrm{T})]$---of pions ($\pi^+$ and $\pi^-$) per recoil charged jet. 
For jets with $R=0.2$ in central Pb+Pb collisions, one can observe clear power-law tails of their constituent pions as in $p+p$ collisions from very low $m_\mathrm{T}$. In sharp contrast, the distribution for jets with $R=0.4$ and $10 < p_\mathrm{T,jet} <20$~GeV in Pb+Pb collisions is much softer, dropping much faster with $m_\mathrm{T}$, and showing a thermal-like feature up to $m_\mathrm{T}\sim 2$~GeV. 

From the thermal fraction of jet energy in Fig.~\ref{fig:3}, the jet shapes in Fig.~\ref{fig:4}, and the constituent transverse mass distributions in Fig.~\ref{fig:5}, it is clear that a significant fraction of energy in jets with large $R$ and intermediate $p_{\rm T,jet}$ still comes from medium response that modifies both the jet shape and constituent momentum distribution. 

{\it \color{blue} Summary and discussions -- }  
Within the LBT model, we found the contribution from jet-induced medium response to the energy of low-$p_\mathrm{T,jet}$ and large-$R$ jets at the partonic level could be much larger than one would expect and ``thermal recoil jets" can emerge from the QGP medium in high-energy heavy-ion collisions. They account for the $p_\mathrm{T,jet}$, $R$, and $\Delta\phi$ dependencies of the enhanced yields of hadron-triggered jets observed in experiments, which challenge the conventional picture of medium modification of hard jets. 

Compared to particles developed from vacuum and medium-modified jet showers within hard jets, particles from medium response are more sparsely distributed in the momentum space. This leads to distinctly different jet shapes between hard and thermal recoil jets: while the shape of the former rapidly decreases with the radial distance from the jet axis, the shape of the latter can increase with the distance. The thermal nature of the jet constituents can be further manifested in their transverse mass distribution. We show that while the $m_\mathrm{T}$ distribution of constituents inside the hard jets resembles that in vacuum, the thermal-like distribution inside the thermal recoil jets is much softer. These characteristics can provide additional signatures of the emergence of thermal recoil jets, though their experimental verification will require dedicated analyses with realistic background treatment. We will extend this study to $\gamma$-jets and jets produced in RHIC experiments in a follow-up paper.

Although thermal recoil jets mainly originate from the QGP medium, they are strongly correlated with the hard jets and qualitatively different from the fake jets formed by uncorrelated particles inside the QGP background. Since the medium response encodes rich information about both jet-QGP interactions and the transport properties of the QGP medium, these thermal recoil jets enable a new approach to probing the color-deconfined nuclear matter.

{\it \color{blue} Acknowledgments -- } We are grateful for valuable discussions with Yaxian Mao. This work was supported by the National Natural Science Foundation of China (NSFC) under Grant Nos.~12475144 (PJ, LY), 14-547 (PJ, YH, LY), 12575146, 12175122, 2021-867, 12321005 (YD, LZ, SC), and 12535010 (XNW).

\bibliographystyle{h-physrev5}
\bibliography{ref}

\end{document}